\documentclass[aps,groupedaddress,a4paper,floatfix,twocolumn,footinbib]{revtex4}
\usepackage{pslatex}
\usepackage{amsfonts}
\usepackage{amssymb}
\usepackage{mathrsfs}
\usepackage{ifthen}
\usepackage[mathscr]{euscript}
\usepackage{graphicx}
\usepackage{fancyhdr}
\usepackage[hypertex=true]{hyperref} %[hyperindex=false] [backref=page]

\begin{document}

\title{
Hot-hole lasers in III-V semiconductors.
}

\author{P. Kinsler}
\email{Dr.Paul.Kinsler@physics.org}
\altaffiliation[New address:]{
Department of Physics, Imperial College,
  Prince Consort Road,
  London SW7 2BW, 
  United Kingdom.
}
\affiliation{
Department of Applied Physics, Technical University Delft, 
Lorentzweg 1, 2628 CJ DELFT, The Netherlands. 
}

\author{W.Th. Wenckebach}
\email{wenckebach@tn.tudelft.nl}
\affiliation{
Department of Applied Physics, Technical University Delft, 
Lorentzweg 1, 2628 CJ DELFT, The Netherlands. 
}

%\renewcommand{\baselinestretch}{1.20}
%\renewcommand{\baselinestretch}{2.00}

%\draft

\lhead{
\includegraphics[height=5mm,angle=0]{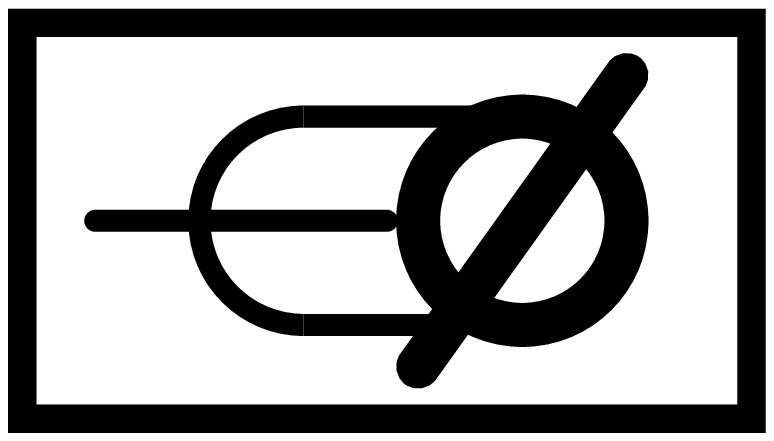}~~
HOTHOLE35}
\chead{~}
\rhead{
\href{mailto:Dr.Paul.Kinsler@physics.org}{Dr.Paul.Kinsler@physics.org}\\
\href{http://www.kinsler.org/physics/}{http://www.kinsler.org/physics/}
%http://www.kinsler.org/physics/}
}

\begin{abstract}

Following the success of p-Ge hot-hole lasers, there is also potential for
using other semiconductor materials, notably III-V's such as GaAs and InSb. 
Previous analysis had suggested that a large effective mass ratio between the
heavy and light holes is advantageous, which implies that InSb would make an
excellent hot-hole laser.  Using our Monte Carlo simulations of both GaAs and
InSb hot-hole lasers in combination with a rate equation model, we see that
previously accepted criteria used to predict performance are not always
reliable, and we suggest suitable alternatives.  The simulation results include
gain and gain bandwidth as a function of field strength and laser frequency,
and alternative field orientations and photon polarizations are considered. 
Comparisons are made with bulk p-Ge systems.  The optimum conditions predicted
by our simulations could then be used in the design of quantum-well hot-hole
lasers.  

\end{abstract}

\pacs{ 42.55.Px , 78.20.Bh , 78.55.Cr , 72.90.+y }

\maketitle
\thispagestyle{fancy}

% 42. Optics (for optical properties of gases, see 51.70; ...
% 42.55.Px  Semiconductor lasers; laser diodes
%
% 78. Optical properties, condensed-matter spectroscopy ...
% 78.20.Bh Theory, models, and numerical simulation
% 78.55.Cr III-V semiconductors
%
% 72. Electronic transport in condensed matter (for ...
% 72.90.+y Other topics in electronic transport in 
%             ... no sub-code for hole-transport-modelling

\newpage

% --------------------------------------------------------------------------

{\em
This paper was published as 
J. Appl. Phys. {\bf 90}, 1692 (2001).
}

\section{Introduction}\label{Sintroduction}

Hot-hole lasers emit in the THz (far-infrared) with an unusually broad gain
spectrum, allowing amplification and generation of laser pulses on a picosecond
time scale
\cite{OQE23-1991,Hovenier-MPSSW-1997apl,Strijbos-etal-1996hcis,Keilmann-T-1992sst,Brundermann-etal-1995ipt,Strijbos-LW-jpcm,Hovenier-KWMPS-1998apl,Murajov-WSPSP-1999apl}.  
This type of laser has been realised in bulk p-doped Germanium (p-Ge), and
produced gains of about 0.25cm$^{-1}$.  Since the THz band has important
potential applications in medical imaging and short-range (e.g.  office)
communications, and due to the widespread applications of GaAs
\cite{Capasso-1990PQED,Shur-1994GaAs,Capasso-SCG-1999pw}, there has been
interest in the potential of III-V materials such as GaAs and InSb as hot-hole
lasers.  Most predictions have been based on simple scattering rate and
effective mass arguments (see e.g. \cite{OQE23-1991}); so to investigate the
possibilities of III-V materials more thoroughly we have done a set of Monte
Carlo simulations for both GaAs and InSb.  We cover a range of field strengths,
orientations, and doping concentrations.

\begin{figure}
\includegraphics[width=80mm,angle=0]{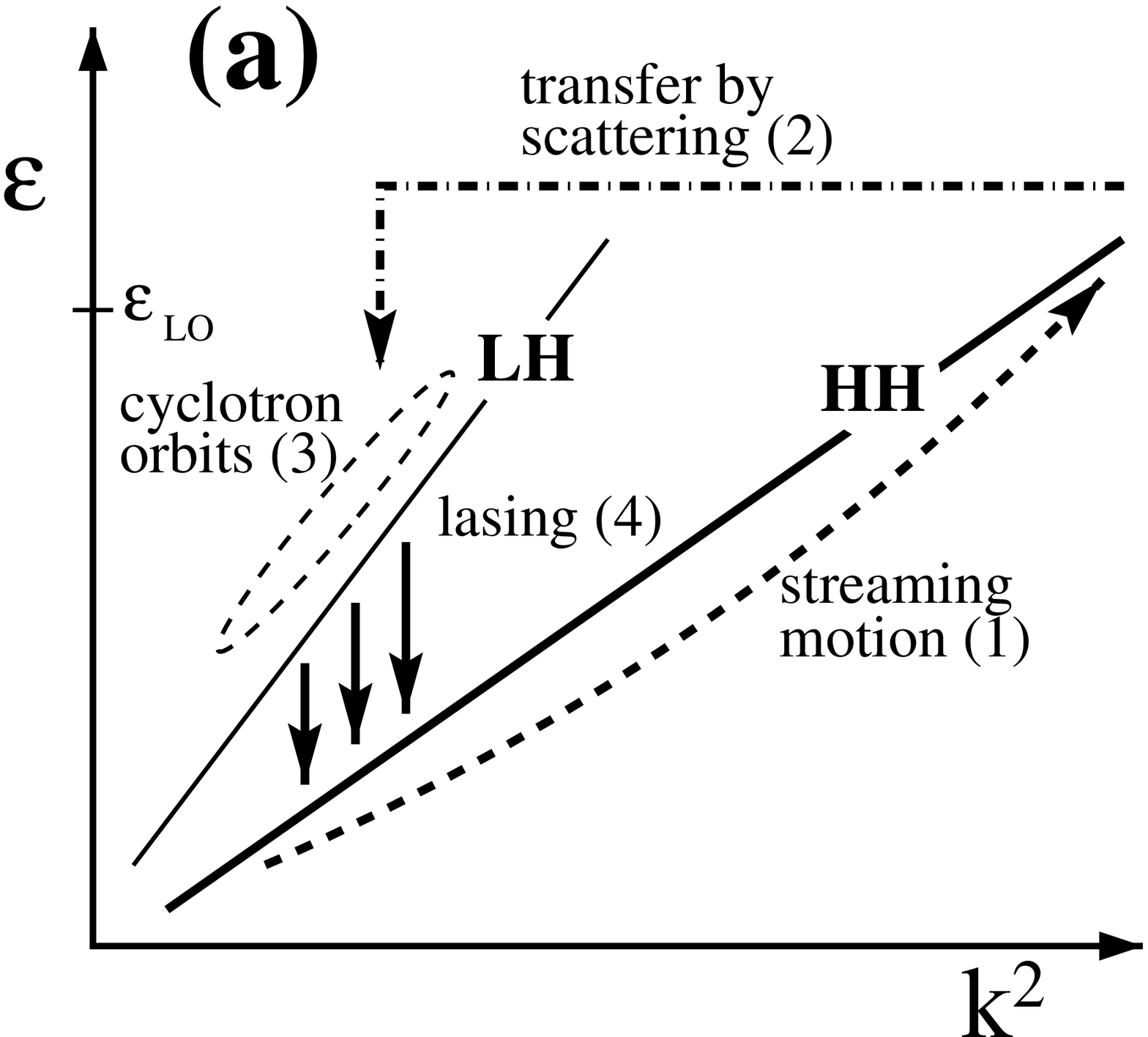}
\includegraphics[width=80mm,angle=0]{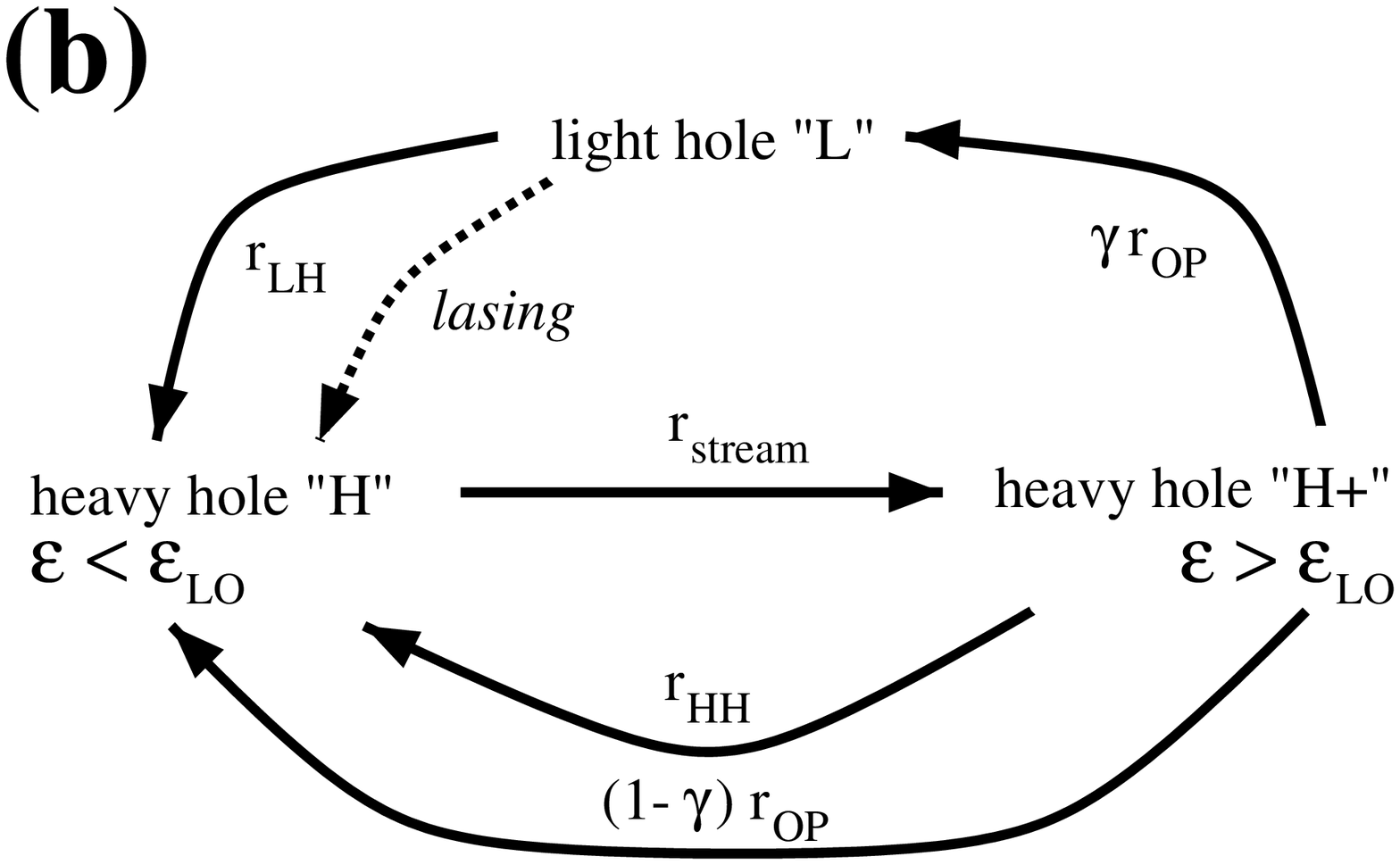}
%\centerline{\epsfig{figure=f1a-lasr.eps,width=80mm,angle=0}}
%\centerline{\epsfig{figure=f1b-rate.eps,width=80mm,angle=0}}
%\includegraphics[height=50mm,angle=0]{laser.eps}
%\epsfig{figure=laser.eps,width=50mm,angle=0}
%\epsfig{figure=laserS.eps,width=50mm,angle=0}
\caption{ 
\label{Flaser} 
%{\bf Flaser: }
The hot-hole laser: (a) shows the streaming motion up the heavy hole band (1), 
the scattering to the light hole band (2), 
the cyclotron orbits in the light hole band (3), 
and the lasing transition (4);
(b) shows a simple three-state model of the operation of the laser, 
with the transition rates being: $r_{stream}$ the rate of streaming
up HH to above the optical phonon energy, $r_{OP}$ the optical phonon emission 
rate, $r_{HH}$ the HH intra-band scattering rate, and $r_{LH}$ the rate of 
depopulation of the LH band.
}
\end{figure}

Hot-hole lasers usually consist of a p-doped Ge crystal in a cryostat cooled to
about 20K, with crossed electric and magnetic fields applied, and are described extensively in Ref. \cite{OQE23-1991}.  In this system,
the ideal lasing cycle for a hole is depicted in Fig. \ref{Flaser} as follows:
(1) the electric field accelerates a heavy hole to above the optical phonon
energy $\epsilon_{LO}$, (2) the hole scatters into the light hole band by
emitting an optical phonon, (3) due to its lighter mass, the hole is now
localised on a closed cyclotron orbit in the light hole band, and this
localisation forms an inversion, and so (4) stimulated emission of a photon
transfers the hole from the light back to the heavy hole band.  In order to get
the streaming motion (1) combined with the cyclotron orbits of (3) it is
necessary to apply the correct ratio of electric to magnetic field strengths as
determined by the heavy to light hole effective mass ratio.  If this is not
done we may get cyclotron orbits which do not reach $\epsilon_{LO}$ in the
heavy hole band, or light holes which reach $\epsilon_{LO}$ and scatter instead
of being localised in cyclotron orbits.  

The results we present in this paper are obtained using a Monte Carlo
simulation (see section \ref{Smontecarlo}) which can deal with for
bandstructure and scattering processes properly.  However, in order to describe
the basic mechanism we now present a simple rate equation model.

The holes are, roughly speaking, in one of three different locations:
{\lq\lq}H{\rq\rq} the lower part of the heavy hole band; {\lq\lq}H+{\rq\rq} the
upper part of the heavy hole band (above $\epsilon_{LO}$); and
{\lq\lq}L{\rq\rq} on cyclotron orbits in the light hole band.  Holes in
{\lq\lq}H{\rq\rq} stream up to {\lq\lq}H+{\rq\rq} with rate $r_{stream}$
\cite{T_LO}, or their streaming is disrupted by intraband scattering with rate
$r_{HH}$.  Holes in {\lq\lq}H+{\rq\rq} will most likely emit an optical phonon,
and either scatter into {\lq\lq}L{\rq\rq} with rate $\gamma r_{OP}$, or back
into {\lq\lq}H{\rq\rq} with rate $\left( 1 - \gamma \right) r_{OP}$.  Holes in
{\lq\lq}L{\rq\rq} can either scatter by phonon or impurity processes back into
{\lq\lq}H{\rq\rq}, or emit a photon and scatter into {\lq\lq}H{\rq\rq} also. 
This is diagrammatically represented on fig. \ref{Flaser}(b), and the
populations ($N_H$, $N_{H+}$, $N_{L}$) in these locations can be described by
rate equations:

\begin{eqnarray}
\frac{d}{dt}N_H &=& - r_{stream} N_H 
                    + \left[ (1-\gamma) r_{OP} + r_{HH} \right] N_{H+}
                    + r_{LH} N_L
,\label{ErateH}
\\
\frac{d}{dt}N_{H+} &=& + r_{stream} N_H 
                       - \left[ r_{OP} + r_{HH} \right] N_{H+}
,\label{ErateHp}
\\
\frac{d}{dt}N_L &=& 
                       + \gamma r_{OP} N_{H+}
                       - r_{LH} N_L
,\label{ErateL}
\end{eqnarray}

where $\gamma$ is the fraction of optical phonon emissions which end up in the
LH band.  Using a density of states ratio argument and $m_{HH} \gg m_{LH}$,
this is $\gamma \approx \left( m_{HH} /m_{LH} \right)^{-3/2}$.  Note that this
model says nothing about the inversion, which is localised in $k$ space.  

None of the steps in the ideal lasing cycle (1-4) are very efficient, and so it
is best to regard the hole transport as being predominantly a sea of unhelpful
scatterings with the lasing cycle superimposed on top of it. It is only if we
can make the lasing cycle strong enough by selecting a suitable material, and
adjusting the field strengths, field directions, or doping concentration will
we see lasing.  Past analysis of the lasing cycle has provided number of
criteria that are usually used to determine whether steps (1-4) will be
sufficiently strong
\cite{Shastin-1991oqe,Komiyama-KHAI-1991oqe,Andronov-1986imw}.  In particular,
criteria for the temperature and electric and magnetic field strengths were
given to ensure that the heavy and light holes would be unlikely to scatter
below the optical phonon energy; that heavy holes do scatter when above this
energy; and that some of these scattered holes reach the light hole band.  They
also favoured a high effective mass ratio $m_{HH}/m_{LH}$, such as that in
InSb, making it easier to ensure that the heavy holes are accelerated to the
optical phonon energy whilst the light holes remain in closed orbits.  We show
here that such arguments, especially the latter, do not always hold true.

% ---------------------------------------------------------------------------
%\end{section}

\section{Simulations}\label{Smontecarlo}

The simulations were done in two stages.  First, for chosen electric ($E$) and
magnetic ($B$) field directions (i.e. $E[01\bar{1}]B[100]$;
$E[112]B[1\bar{1}0]$; $E[2\bar{1}\bar{1}]B[0\bar{1}1]$;
$E[\bar{1}\bar{1}1]B[112]$ -- see \cite{Lok-thesis-Ge} ), one Monte Carlo
simulation was done for each point of a grid of field strengths covering the
range of interest.  The simulations produced distribution functions of heavy
and light holes and the average scattering times for the system under the
chosen conditions.  The second stage used the distributions to calculate
inversion-gain cross-sections $\sigma_I$; and the average scattering times were
used in a Drude model to calculate the optical absorption $\sigma_a$.  These
two were then combined and multiplied by the impurity concentration $n_i$ to
give the net gain per unit length $\lambda = ( \sigma_I - \sigma_a ) n_i$ as a
function of photon frequency for each grid point.  This three dimensional
dataset of net gains was then analyzed to produce contour plots of maximum
gain, enabling us to see the best directions and strengths of $E$ and $B$ to
choose to get optimum gain.  

Our Monte Carlo simulations\cite{MonteCarlo} of this system include a full
$k.p$ band structure calculation \cite{Kane-1966ss} and all important
scattering processes: optical phonon scattering due to deformation potential
(OPD) and polar interactions (OPP), acoustic phonon scattering due to
deformation potential (ACD) and piezoelectric interactions (ACP), and
(Coulomb) ionized impurity (IIM) scattering \cite{Ridley-QPS,Reggiani-1985hts}.
We treat the effect of the electric and magnetic fields classically, giving the
holes continuous trajectories in $k$-space -- comparison with experiment for
p-Ge systems indicates that this approximate treatment is adequate for the
field strengths we consider.  All simulations are done at a lattice temperature
of 20K.  They follow the progress of a single hole through a large number of
scatterings (typically $\sim 16000$), with the ergodic theorem being used to
justify the use of the simulation's time-average as an ensemble average.  Since
both the hole-hole scattering processes and ionized impurity scatterings are
mediated by screened coulomb potentials,  this means that hole-hole scattering
can be allowed for by doubling the impurity concentration in the simulation --
since the hole concentration is equal to the impurity concentration
\cite{Andronov-1986book}.  

The Monte Carlo simulations used the standard overestimation technique where
for each scattering process the post-scattering direction of the hole was
chosen at random, and the differential scattering rate was overestimated by an
isotropic rate just higher than its maximum value.  However, this proved to be
very slow for the ionised impurity scattering, especially at low impurity
concentrations.  Therefore, in this case we weighted the choice of scattering
angle by the angular dependence of the scattering rate, and thus avoided
generating a large proportion of inefficient overestimations.

Optical absorption was estimated using a Drude free carrier model, which uses
zone centre effective masses and the hole scattering times calculated in the
Monte Carlo simulations.  This model assumes that each scattering completely
randomises the phase of the particle.  However, IIM
scattering, which makes a significant contribution to the average hole
scattering times, involves a large proportion of scatterings which only
slightly deflect the hole.  As a result, the ordinary treatment of IIM
scattering would significantly overestimate the scattering rate used to
calculate the optical absorption.  We avoid this problem because our IIM
algorithm weights the choice of angle, while the scattering rate itself is
isotropic.  By choosing an algorithm compatible with our model of optical
absorption, we achieved both an efficient model of IIM scattering whilst
avoiding the necessity of a more complex absorption model.

% ---------------------------------------------------------------------------
%\end{section}

\section{Germanium}\label{Sgeresults}

\begin{figure}
\includegraphics[width=80mm,angle=0]{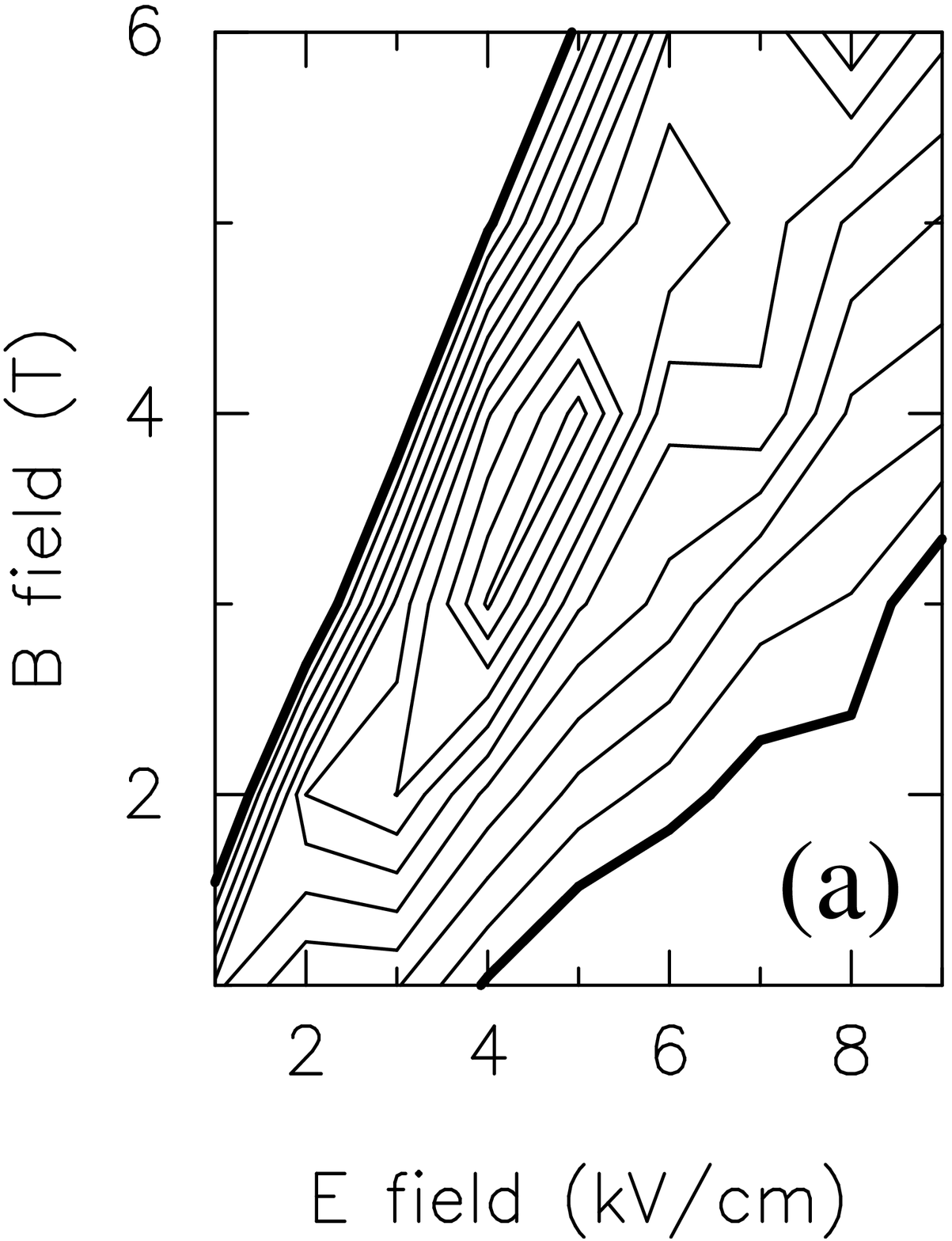}
\includegraphics[width=80mm,angle=0]{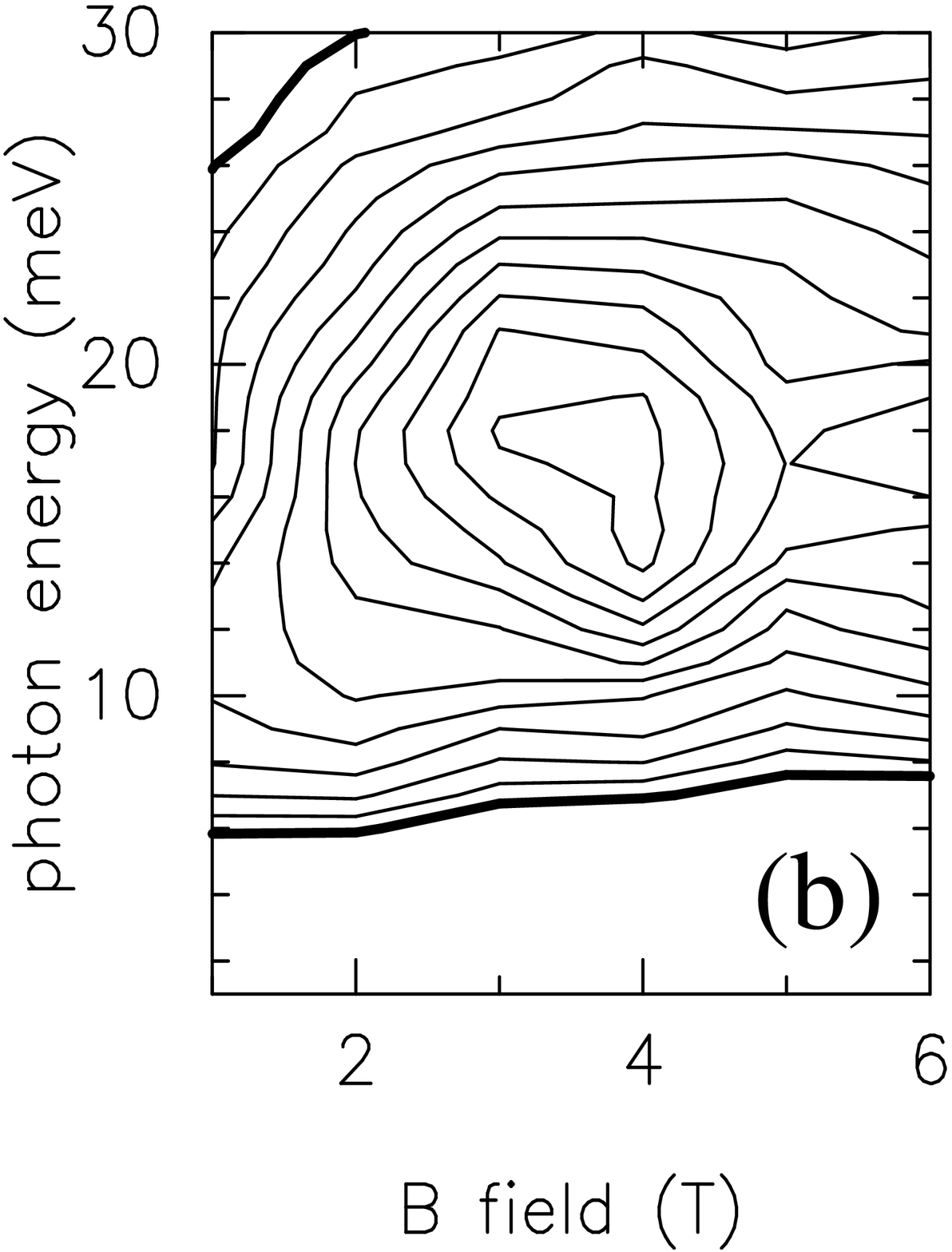}
%\centerline{\epsfig{figure=f2a-GeBE.eps,width=80mm,angle=0}}
%\centerline{\epsfig{figure=f2b-GefB.eps,width=80mm,angle=0}}
%\epsfig{figure=opp-dk.eps,width=50mm,angle=0}
\caption{ 
\label{Fge-005} 
%{\bf Fge-005: }
Gain in p-Ge, $n_i= 0.025 \times 10^{16}$cm$^{-3}$, laser photon 
polarization in the $\vec{E}$ direction.
(a) The net gain after 
free-carrier optical absorption has been taken into account. The thick
contour is at zero, and contours step up in gain 
by 0.025cm$^{-1}$. 
(b) The net gain as a function of magnetic field and photon frequency 
(for the optimum  electric field).
}
\end{figure}

To validate our model and simulation code, we first did some simulations of
hot-hole lasers in p-Ge and compared our results against those typically given
in the literature.  The optimum gain contour plots are shown on fig. 
\ref{Fge-005}, and table \ref{Ge-table} shows the relative frequencies of the
different scattering types at the point of optimum gain. The columns ACD, IIM,
and OPD on this table refer to the acoustic phonon, ionised impurity, and
deformation potential optical phonon scattering processes respectively.  These
simulations also provided us with a good reference with which our
GaAs and InSb simulation results can be to compared.  The results, for an
impurity concentration of $n_i = 0.025 \times 10^{16}$cm$^{-3}$, are shown in
Fig.  \ref{Fge-005}, where the optimum gain is about $0.25$/cm at $F = 5$kV/cm
$[2\bar{1}\bar{1}]$ and $B = 4$T $[0\bar{1}1]$.  

% (c) The optimum net gain as a function of electric field and frequency (for
% all magnetic fields) (d) The bandwidth of the gain, where contours are spaced
% by 20meV ($\sim$4THz), and indicate the energy range over which net gain is
% predicted. 
% (e) The frequency at the best gain, with contours
% are spaced by 1meV.

%\epsfig{figure=Ge-net-E-BEv.ps,height=55mm,angle=0}
%\epsfig{figure=Ge-net-E-fBv.ps,height=55mm,angle=0}

%%  %\epsfig{figure=Ge-net-E-fEv.ps,height=55mm,angle=0}
%%  %\epsfig{figure=Ge-net-E-BEw.ps,height=55mm,angle=0}
%%  %\epsfig{figure=Ge-net-E-BEc.ps,height=55mm,angle=0}

% MAT=Ge
% MAT=GaAs
% MAT=InSb
% DIRN=E
% for II in d3-net-${DIRN}-BE?.ps d3-net-${DIRN}-fEv.ps d3-net-${DIRN}-fBv.ps; do cp -a ${II} ~/_work/_current/_HHBULK/${MAT}${II#d3}; done

\begin{table}[h]
\caption{The relative frequencies of different scattering
types in Ge at a point of optimum gain: $E=5$kV/cm $[2\bar{1}\bar{1}]$, 
$B=4$T $[0\bar{1}1]$ and $n_i=0.025 \times 10^{16}$.  
There were a total of 16000 scatterings, and the hole was in 
the LH band for $250 \pm 44 ps$, and the HH band for $1076 \pm 49$ ps. 
The average time between collisions in the LH and HH bands were 
$t_{LH}=4.9$ps and $t_{HH}=1.1$ps respectively.
NB: $m_{HH}/m_{LH}\approx 8$; 
$r_{stream} \approx 1.3$ps$^{-1}$;
}
\begin{center}
\begin{tabular}{ c c c c c }
Transition&            ACD & IIM &   OPD & Total  \\ \hline
HH $\rightarrow$ HH &  749 & 161 & 13619 & 14529  \\
HH $\rightarrow$ LH &   48 &   8 &   612 &   676  \\
LH $\rightarrow$ LH &    4 &  95 &    21 &   120  \\
LH $\rightarrow$ HH &  158 &  90 &   428 &   676
\end{tabular}
\end{center}
\label{Ge-table}
\end{table}

% LH band scattering time $t_{LH}=4.9$ps
% HH band scattering time $t_{HH}=1.1$ps
%  T_{sim-L} =  250 \pm 44 ps
%  T_{sim-H} = 1076 \pm 49 ps
%  \gamma \sim 0.044
%  r_{LH} \approx \frac{120}      { 250ps}   \approx  0.48 /ps
%  r_{HH} \approx \frac{749+161}  {1076ps}   \approx  0.85 /ps
%  r_{OP} \approx \frac{13619+612}{1076ps}   \approx 13.2  /ps
%
%; $1/T_{c,LH} = f_{c,LH} \approx 4$THz at 1T.

These simulations also enabled us to test the accuracy of our simple model of
optical absorption against the more sophisticated model of Lok
\cite{Lok-thesis-Ge,Lok-thesis-values}, which included second order transitions
of phonons and photons, or ionised impurities and photons; while all scattering
processes where taken into account.  We obtained optical absorptions within
$\sim 25$\% of Lok at photon energies of 5meV -- due to the inverse dependence
of absorption with frequency, the fit was much better above 5meV, but worse
below.  Lok's model has optical phonon scattering dominating optical
absorption, with IIM scattering appearing to have a negligible effect.  In his
results, the lower gains at higher impurity concentrations are due to IIM
scattering from the light hole to the heavy hole band, causing a loss of
inversion.  This has a more significant effect on the optical gain than the
extra optical absorption caused by the IIM scattering. Using our algorithm that
weights the choice of IIM scattering angle, we see the same.

% ---------------------------------------------------------------------------
%\end{section}

\section{Gallium Arsenide}\label{Sgaasresults}

The obvious III-V material to investigate is GaAs, as it is already widely used
for many other applications
\cite{Capasso-1990PQED,Shur-1994GaAs,Capasso-SCG-1999pw}.  It has zone center
effective masses of roughly $m_{HH} \approx 0.50$ and $m_{LH} \approx 0.07$,
with a ratio rather similar to that in germanium (Ge).  The main difference
compared to Ge is that GaAs has additional scattering processes -- it is a
polar material, so there are also piezo-electric phonon (ACP) processes and
polar optical phonon (OPP) processes to consider.  While the piezo-electric
processes are usually not very significant, the polar optical phonons have a
significant role, largely because their strength is inversely dependent on
the phonon momentum.

%  (see Fig. \ref{Foppdk}).

\begin{figure}
\includegraphics[width=80mm,angle=0]{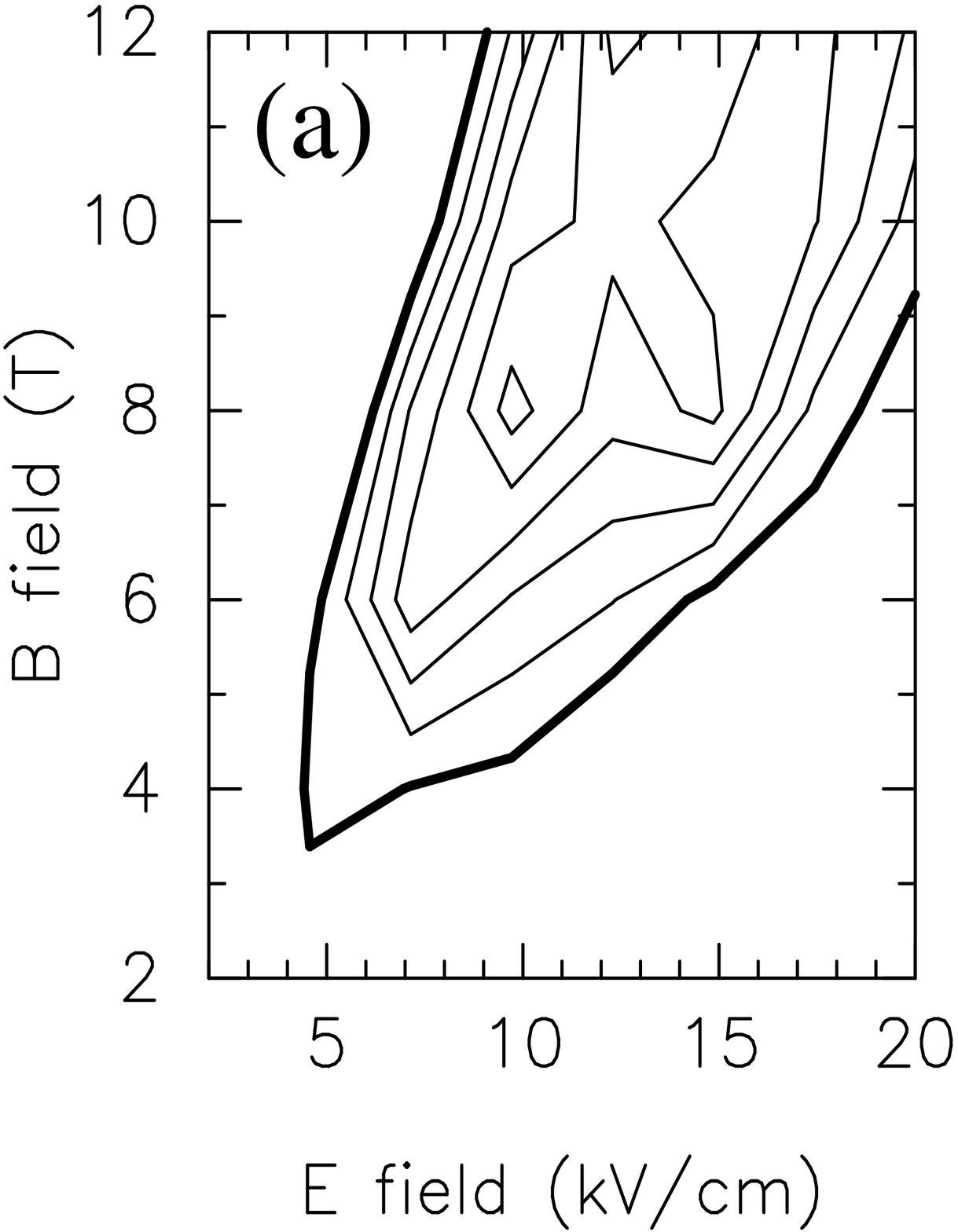}
\includegraphics[width=80mm,angle=0]{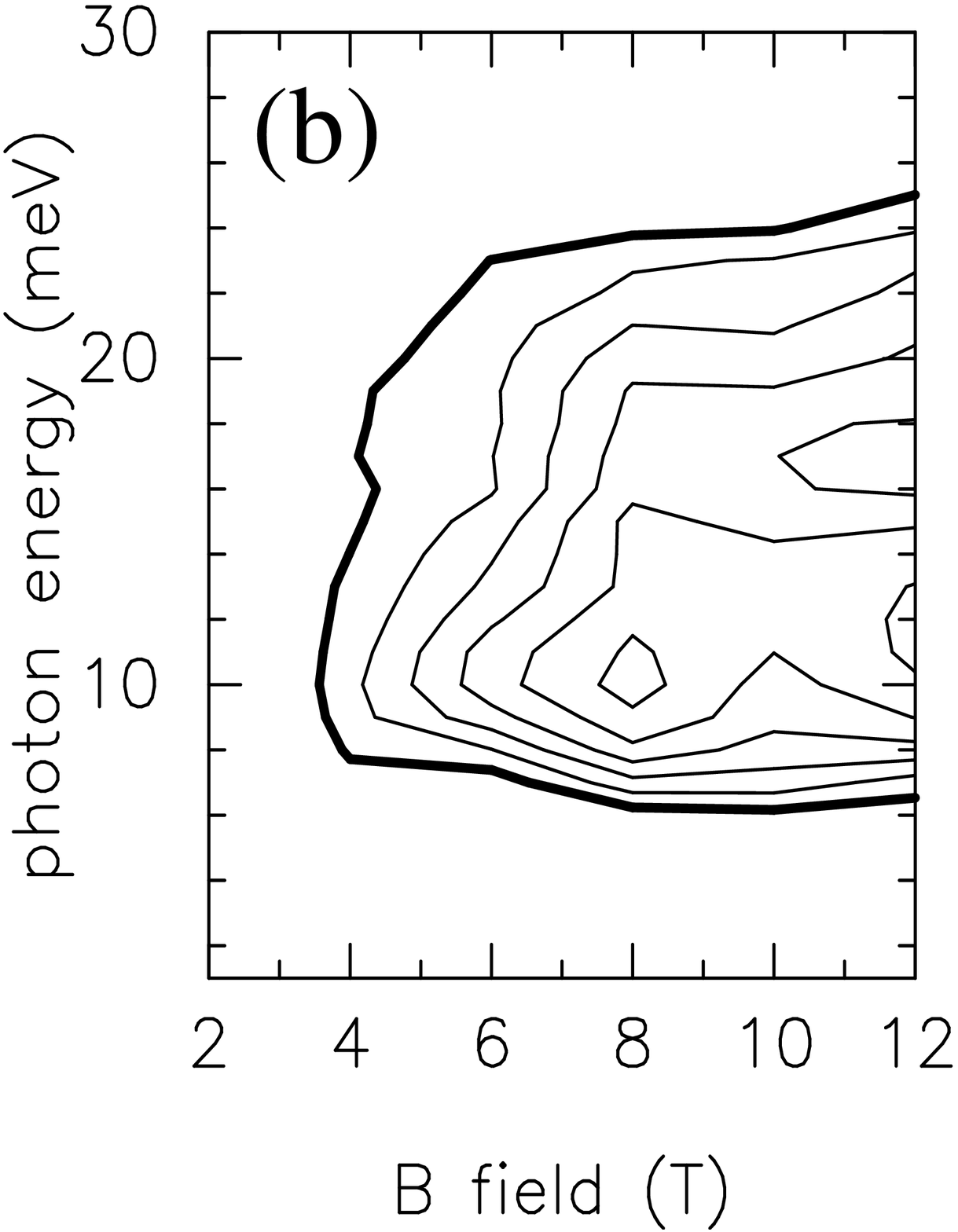}
%X%\centerline{\epsfig{figure=GaAs-net-E-BEv.eps,width=140mm,angle=0}}
%X%\centerline{\epsfig{figure=GaAs-net-E-fBv.eps,width=140mm,angle=0}}
\caption{ 
\label{Fgaas-005} 
%{\bf Fgaas-005: }
Gain in p-GaAs, $n_i= 0.025 \times 10^{16}$cm$^{-3}$, laser photon 
polarization in the $\vec{E}$ direction.
(a) The net gain after 
free-carrier optical absorption has been taken into account. The thick
contour is at zero, and contours step up in gain 
by 0.025cm$^{-1}$. 
(b) The net gain as a function of magnetic field and photon frequency 
(for the optimum  electric field).
}
\end{figure}

% (c) The optimum net gain as a function of electric field and frequency (for
% all magnetic fields) (d) The bandwidth of the gain, where contours are spaced
% by 20meV ($\sim$4THz), and indicate the energy range over which net gain is
% predicted.  
% (e) The frequency at the best gain, with contours
% are spaced by 1meV.

%\epsfig{figure=GaAs-net-E-BEv.ps,height=55mm,angle=0}
%\epsfig{figure=GaAs-net-E-fBv.ps,height=55mm,angle=0}

%%  %\epsfig{figure=GaAs-net-E-fEv.ps,height=55mm,angle=0}
%%  %\epsfig{figure=GaAs-net-E-BEw.ps,height=55mm,angle=0}
%%  %\epsfig{figure=GaAs-net-E-BEc.ps,height=55mm,angle=0}

%%  % Top to bottom the laser photon polarizations are in the 
%%  % $\vec{E}$, $\vec{B}$ and $\vec{E}\times \vec{B}$ directions. 

\begin{table}[h]
\caption{The relative frequencies of different scattering
types in GaAs at a point of optimum gain: $E=8$kV/cm $[2\bar{1}\bar{1}]$, 
$B=7$T $[0\bar{1}1]$
and $n_i=0.025 \times 10^{16}$.  
There were a total of 16000 scatterings, and the hole was in 
the LH band for $180 \pm 25 ps$, and the HH band for $687 \pm 5 ps$. 
The average time between collisions in the LH and HH bands were 
$t_{LH}=3.8$ps and $t_{HH}=0.7$ps respectively.
NB: $m_{HH}/m_{LH}\approx 8$; 
$r_{stream} \approx 1.7$ps$^{-1}$
}
\begin{center}
\begin{tabular}{ c c c c c c c }
Transition&            ACD & ACP & IIM &  OPD & OPP   & Total \\ \hline
HH $\rightarrow$ HH &  899 & 186 & 383 & 4198 & 9001  & 14667  \\
HH $\rightarrow$ LH &   43 &   4 &  34 &  188 &  297  &   566  \\
LH $\rightarrow$ LH &    3 &   7 & 181 &    1 &    9  &   201  \\
LH $\rightarrow$ HH &  222 &  41 & 193 &    8 &  102  &   566
\end{tabular}
\end{center}
\label{GaAs-table}
\end{table}

% LH band scattering time $t_{LH}=3.8$ps
% HH band scattering time $t_{HH}=0.7$ps
%  T_{sim-L} =  180 \pm 25 ps
%  T_{sim-H} =  687 \pm  5 ps
%  \gamma \sim 0.044
%  r_{LH} \approx \frac{201}              { 180ps}   \approx  1.1 /ps
%  r_{HH} \approx \frac{899+186+383}      { 687ps}   \approx  2.1 /ps
%  r_{OP} \approx \frac{4198+9001+188+297}{ 687ps}   \approx 19.9 /ps
%
%; $1/T_{c,LH} = f_{c,LH} \approx 2$THz at 1T.

Simulations were done for a range of electric and magnetic field magnitudes,
from 2kV/cm to 20kV/cm, and 2T to 12T.  The maximum fields were chosen somewhat
higher than is usually practicable in experiment to ensure we covered the full
range of interest.  For $n_i = 0.025 \times 10^{16}$cm$^{-3}$, the $E/B$ ratio
for optimum gain was $1.14$ (cf $1.25$ for Ge).

A full array of simulations was done for $n_i = 0.025 \times
10^{16}$cm$^{-3}$, an electric field direction of $E[2\bar{1}\bar{1}]$, and a
magnetic field direction of $B[0\bar{1}1]$ (relative to the crystal axes).  This
moderate impurity concentration showed a best net gain of $0.08$cm$^{-1}$ at
high fields ($\sim$8kV/cm, 7T) (as shown in Fig. \ref{Fgaas-005}).  Although
reasonable inversion-gain occurs down to fields of a few kV/cm (or T), the
optical absorption is strong enough to spoil it except at higher fields. 
The net gain for a range of different field orientations were also checked at
the optimum 8kV/cm, 7T point, but $E[2\bar{1}\bar{1}]$ \& $B[0\bar{1}1]$ were
confirmed as giving the greatest gain.  In general, the orientation dependence
can be significant, with sometimes up to a factor of two difference in the net
gain.

Next, simulations were done at higher and lower impurity concentrations for the
best-gain value of $E$ and $B$ fields. The higher impurity concentration ($n_i
= 0.100 \times 10^{16}$cm$^{-3}$) results indicated that there is at best a
marginal gain.  This is because the increased impurity scattering has two
effects.  First, the extra scattering increases the calculated optical
absorption; and secondly, impurity scattering is more efficient at depopulating
the LH band by producing LH $\rightarrow$ HH scattering than it is at
repopulating it with HH $\rightarrow$ LH scatterings.  A table of the relative
frequencies of the different scattering types at $E = 8$ kV/cm
$[2\bar{1}\bar{1}]$, $B = 7$ T $[0\bar{1}1]$ is given in Table \ref{GaAs-table}.

The lower impurity concentration simulations ($n_i = 0.005 \times
10^{16}$cm$^{-3}$) show a better net-gain cross-section $\sigma_I - \sigma_A$. 
However, since the number of holes in the simulation is related to the impurity
(doping) concentration, fewer impurities also means fewer holes, and fewer
holes means that the gain per length $\lambda$ is proportionally reduced.  So
although the net-gain cross-section is up to double that for $n_i = 0.025
\times 10^{16}$cm$^{-3}$, the net-gain per cm reduces to $\lambda \sim
0.04$cm$^{-1}$.

The dependence of the net gain on the polarization of the output light is
small.  Typically one orientation (of $B$, $E$, or $E \times B$) is better than
the other two, which are otherwise very similar.  Output light polarized in the
$B$ direction requires the hot-hole laser to be in the Voigt configuration, but
$E$ polarized output can be achieved by either the Voigt or Faraday
configurations.  In our simulations, the polarization in the electric field
($E$) direction gave the best gain; this contrasts with the comments in Ref. 
\cite{Shastin-1991oqe} saying that a polarization parallel to the magnetic
field ($B$) minimises heavy hole scattering and hence the optical absorption.

% ---------------------------------------------------------------------------
%\end{section}

\section{Indium Antimonide}\label{Sinsbresults}

An often mentioned candidate for a III-V hot-hole laser is p-InSb, because its
effective mass ratio $m_{HH}/m_{LH}$ is much greater than that in either GaAs
or Ge.  The simple argument is that since the light hole cyclotron orbits can
be more tightly confined in $k$-space, this leads to a more pronounced
inversion.  However, since our simulation code cannot treat light hole Landau
levels, we restricted our simulations to magnetic fields under 2T; and so were
unable to make predictions at the higher fields where the confined
light hole cyclotron orbits might give a significantly increased gain.

\begin{figure}
\includegraphics[width=80mm,angle=0]{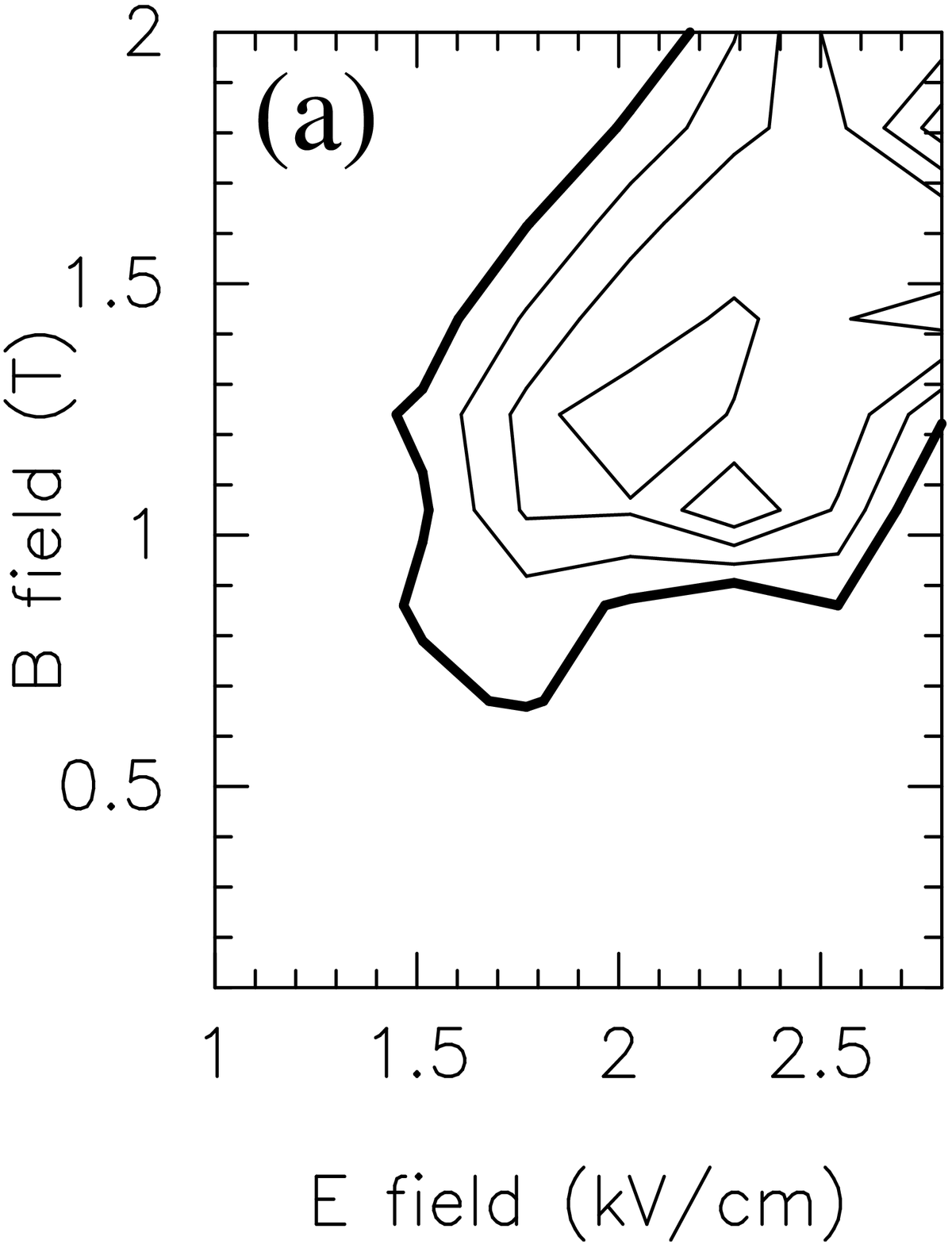}
\includegraphics[width=80mm,angle=0]{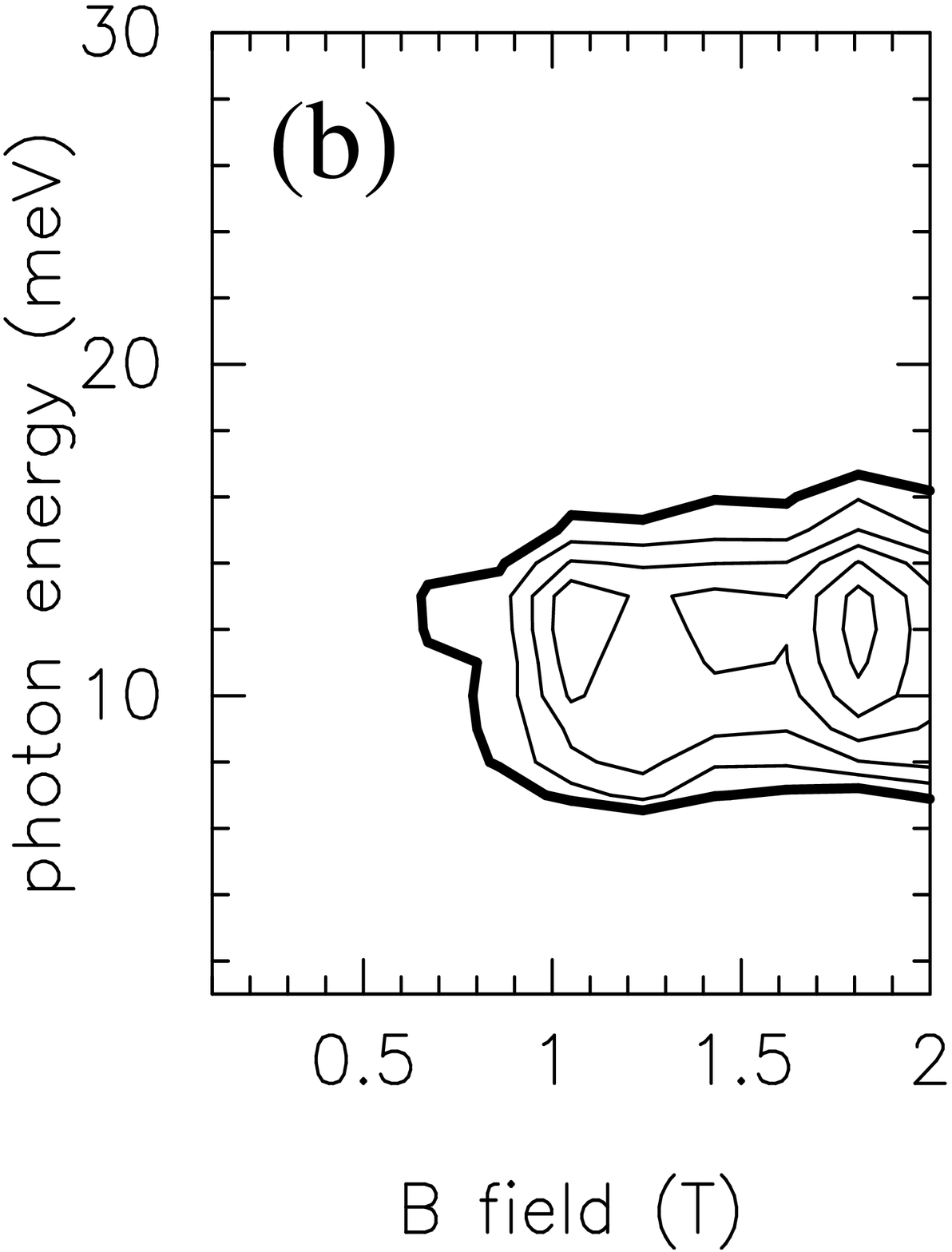}
%\centerline{\epsfig{figure=f4a-InSb.eps,width=80mm,angle=0}}
%\centerline{\epsfig{figure=f4b-InSb.eps,width=80mm,angle=0}}
%X%\centerline{\epsfig{figure=InSb-net-E-BEv.eps,width=140mm,angle=0}}
%X%\centerline{\epsfig{figure=InSb-net-E-fBv.eps,width=140mm,angle=0}}
\caption{ 
\label{Finsb-005} 
%{\bf Finsb-005: }
Gain in p-InSb, $n_i= 0.025 \times 10^{16}$cm$^{-3}$, laser photon
polarization in the $\vec{E}$ direction.
(a) The net gain after 
free-carrier optical absorption has been taken into account. The thick
contour is at zero, and contours step up in gain 
by 0.025cm$^{-1}$. 
(b) The net gain as a function of magnetic field and photon frequency 
(for the optimum  electric field).
}
\end{figure}

% (c) The optimum net gain as a function of electric field and frequency (for
% all magnetic fields) (d) The bandwidth of the gain, where contours are spaced
% by 20meV ($\sim$4THz), and indicate the energy range over which net gain is
% predicted.  
% (e) The frequency at the best gain, with contours are spaced by 1meV.  }

%\epsfig{figure=InSb-net-E-BEv.ps,height=55mm,angle=0}
%\epsfig{figure=InSb-net-E-fBv.ps,height=55mm,angle=0}

%%  %\epsfig{figure=InSb-net-E-fEv.ps,height=55mm,angle=0}
%%  %\epsfig{figure=InSb-net-E-BEw.ps,height=55mm,angle=0}
%%  %\epsfig{figure=InSb-net-E-BEc.ps,height=55mm,angle=0}

\begin{table}[h]
\caption{The relative frequencies of different scattering
types in InSb at a point of optimum gain: $E=1.8$kV/cm $[01\bar{1}]$, 
$B=1$T $[100]$
and $n_i=0.025 \times 10^{16}$.  
There were a total of 16000 scatterings, and the hole was in 
the LH band for $55 \pm 17 ps$, and the HH band for $1326 \pm 13$ ps. 
The average time between collisions in the LH and HH bands were 
$t_{LH}=4.5$ps and $t_{HH}=1.4$ps respectively.
NB: $m_{HH}/m_{LH}\approx 30$;
$r_{stream} \approx 0.55$ps$^{-1}$
}
\begin{center}
\begin{tabular}{ c c c c c c c }
Transition&            ACD & ACP & IIM &  OPD & OPP   & Total \\ \hline
HH $\rightarrow$ HH & 1681 &  47 & 648 & 3610 & 9679  & 15665  \\
HH $\rightarrow$ LH &   15 &   0 &  16 &   47 &   70  &   148  \\
LH $\rightarrow$ LH &    2 &   0 &  37 &    0 &    0  &    39  \\
LH $\rightarrow$ HH &   50 &   1 &  50 &    3 &   44  &   148
\end{tabular}
\end{center}
\label{InSb-table}
\end{table}

% LH band scattering time $t_{LH}=4.5$ps
% HH band scattering time $t_{HH}=1.4$ps
%  T_{sim-L} =   55 \pm 17 ps
%  T_{sim-H} = 1326 \pm 13 ps
%  \gamma \sim 0.044
%  r_{LH} \approx \frac{39}             {  55ps}   \approx  0.7 /ps
%  r_{HH} \approx \frac{1681+47+648}    {1326ps}   \approx  1.8 /ps
%  r_{OP} \approx \frac{3610+9679+47+70}{1326ps}   \approx 10.1 /ps
%
%; $1/T_{c,LH} = f_{c,LH} \approx 2THz$ at 1T.

From Fig. \ref{Finsb-005} we can see that the peak gain at $n_i = 0.025 \times
10^{16}$cm$^{-3}$ is about 0.05cm$^{-1}$, and the trend is the usual one for
better inversion at higher fields.  The dependence of the net gain on the
polarization of the output light is small, as for GaAs, and indeed Ge.  For
InSb, the best field orientation is not an electric field direction of
$[2\bar{1}\bar{1}]$, and a magnetic field direction of $[0\bar{1}1]$; but
$E[01\bar{1}]$ $B[100]$.  A table of the relative frequencies of the different
scattering types at $E = 1.8$ kV/cm $[0\bar{1}1]$, $B = 1.0$ T $[100]$ is given
in Table \ref{InSb-table}.

As with GaAs, simulations were done for InSb at the same higher and lower
impurity concentrations ($n_i = 0.100 , 0.005  \times 10^{16}$cm$^{-3}$) for
the best-gain value of electric and magnetic fields. The higher impurity
concentration produced a prediction of no net gain -- for the same reasons as
when the higher impurity concentration reduced the gain in GaAs.  The low
impurity concentration gave reduced gain, with the resulting smaller hole
concentration again overcoming the slight increase in gain cross section.

% ---------------------------------------------------------------------------
%\end{section}

\section{Discussion}\label{Sconclusion}

These results indicate a number of important considerations for hot-hole laser
operation in bulk III-V materials.  Firstly, control of the impurity
concentration is critical, as e.g. in GaAs, the material can go from no
net-gain through its peak back to no net-gain with only a factor of 20 or so
range in impurity concentration.  Secondly, as in Ge, the orientation of the
electric and magnetic fields with respect to the crystallographic axes is
important -- while the difference in the amount of net gain from the best
choice in InSb to the second best choice is not large, the region over which it
can be obtained is.  This then is particularly important if there are
restrictions on the maximum fields allowed.  Finally, it is clear that the
performance in p-GaAs and p-InSb is not as good as in p-Ge hot-hole lasers; the
main reason being that the non-polar p-Ge does not have the polar optical
phonon (OPP) and acoustic pizeoelectric phonon (ACP) scattering processes, and
that this extra scattering has a more significant negative effect than the
benefits of the improved effective mass ratio.

For GaAs, OPP scattering transfers more holes into the LH band than deformation
optical phonon (OPD) scattering -- however, it also scatters more out.  As a
result, the net transfer of holes into the LH band is similar for both optical
phonon processes.  The much greater LH $\rightarrow$ HH rate for OPP is due to
the small momentum transfer $\Delta k$ in such processes: the OPP rate is
inversely proportional to $\Delta k$.  Also note that the GaAs ionised impurity
scattering (IIM) rates are roughly double those for Ge -- this is largely due
to the difference in dielectric constants, which appear to the fourth power in
the denominator of the scattering rate prefactor.  The dielectric constant for
Ge is about 16, compared to 13 for GaAs.  Also, the dielectric constant affects
the screening length which also alters the IIM scattering rates.  Unfortuately
it is not possible to easily compare the IIM scattering rates and dielectric
constants between Ge or GaAs and InSb, as their effective mass ratios are so
different -- $\left(m_{HH}/m_{LH}\right)_{Ge, GaAs} \approx 8$, but
$\left(m_{HH}/m_{LH}\right)_{InSb} \approx 30$.

In InSb the net transfer of holes into the LH band is similar for
both optical phonon processes.  Also the HH $\rightarrow$ LH scattering rates
in InSb are consistently lower than for GaAs, which is due largely to the
lower LH density of states caused by InSb's lower LH effective mass.

Note that for both GaAs and InSb in the above discussion, a large effective
mass ratio $m_{HH}/m_{LH}$ was either not mentioned as important (GaAs), or was
regarded as bad (InSb).  This is in contrast to simple arguments which state
that large ratio is good because it helps ensure efficient streaming up the HH
band whilst retaining good cyclotron orbits in the LH band -- unfortunately,
too large a ratio brings the LH density of states down too low, inhibiting
population transfer to the LH band and hence inhibiting inversion. The simple
arguments also miss the influential role of the dielectric constant in
controlling impurity scattering rates (GaAs).

Next, we need to consider the criteria given in past analyses as described in
section \ref{Sintroduction}.  Here we can clarify their roles using the simple
rate-equation model (Eqs. \ref{ErateH}--\ref{ErateL}) described earlier. For
$N_L \ll N_H + N_{H+}$, the fraction of population in the LH band in the steady
state is

\begin{eqnarray}
P 
= 
\frac{N_L}{N_H + N_{H+}}
= 
  \frac{\gamma r_{OP} r_{stream}}
       { r_{LH} \left( r_{OP} + r_{stream} + r_{HH} \right) } 
.
\label{NLoNH}
\end{eqnarray}

The three important criteria are: (i) a large $\gamma$, which means a small
$m_{HH}/m_{LH}$ ratio (but not so small as to disrupt either the HH streaming
or LH cyclotron orbits); (ii) that $r_{HH}$ should be as small as possible
compared to $r_{stream}+r_{OP}$; and (iii) that $r_{LH}$ should be much less
than the smaller of $r_{stream}$ or $r_{OP}$.  We have extracted these
scattering rates from our optimum gain Monte Carlo simulations, and using these
in eqn.(\ref{NLoNH}) gives the same trends as is observed for the net gain --
$P = 0.10$, $0.08$, $0.05$ for the Ge, GaAs, and InSb simulations respectively.
Also, the simulation $\gamma$'s were generally close to the ratio of
effective masses.

In summary, we have presented a rate equation model of hot-hole lasers and
simulation results that clarify the roles of performance
criteria presented in earlier analyses.  We show that in general
III-V materials should have an acceptable performance, which contrasts with
previous criteria that suggested more optimistically that (e.g.) InSb should
make an excellent hot-hole laser.  This demonstrates the need to treat the
lasing cycle as a whole, as in our rate equation model, rather than treating
each step in isolation.  Finally, the performance of III-V
hot-hole lasers might most easily be improved by moving to a modulation doped
quantum wells -- thus eliminating the  impurity scattering by moving the
impurities away from the active region, and giving us opportunity to engineer
the band structure.

% ---------------------------------------------------------------------------
%\end{section}

\section*{Acknowledgements}
\label{acknowledgements}

\vspace{2mm}
\noindent

We would like to thank C.A.M. Haarman for implementing polar
phonon scattering in the simulation code.
This work is funded by the European Commission via the 
program for Training and Mobility of Researchers.

% ---------------------------------------------------------------------------
%\end{section}

% ---------------------------------------------------------------------------


\begin{thebibliography}{99}
\bibitem{T_LO}
The time take to get from the bottom of the heavy hole band to $\epsilon_{LO}$
is $t_{stream} = \left( \sqrt{ m^* \epsilon_{LO}} / E \right) \times 1.07$ (ps)
where $m^*$ is relative to the electron mass, $\epsilon_{LO}$ is in meV, and
$E$ is the electric field in kV/cm; $r_{stream} = 1/t_{stream}$.  If it does
not scatter, it will return to low energy by completing a cyclotron orbit in a
time $T_{cH} = m_{HH}/eB$.
\bibitem{Lok-thesis-values}
Absorption was compared for for $E=2$kV/cm $[1\bar{1}0]$, 
$B=1.3$T $[\bar{1}\bar{1}1]$, temperature 10K, $n_i=1.3\times10^{14}$cm$^{-1}$.
\end{thebibliography}
\end{document}